\documentclass{ws-mpla}

\newcommand{\be}{\begin{equation}}
\newcommand{\ee}{\end{equation}}
\newcommand{\bea}{\begin{eqnarray}}
\newcommand{\eea}{\end{eqnarray}}

\def\ket#1{\hbox{$\vert #1\rangle$}}   % definition of ket
\def\onet{{\textstyle {1\over 3}}}

\def\oneseven{{\textstyle {1\over 7}}}

\def          % circa \ge
\simeq{{\ \lower2pt\hbox{$-$}\mkern-13mu \raise2pt \hbox{$\sim$}\ }}

\def           % per gamma Dirac
\slasha#1{{\hbox{$\not$} \mkern-3mu\hbox{$\,#1\,$}}}

\begin{document}

\markboth{S. Boffi and B. Pasquini}
{Probing the parton content of the nucleon}

\catchline{}{}{}{}{}
%
%%%%%%%%%%%%%%%%%%%%%%%%%%%%%%%%%%%%%%%%%%%%%%%%%%%

\title{PROBING THE PARTON CONTENT OF THE NUCLEON}

\author{\footnotesize SIGFRIDO BOFFI\footnote{
Sigfrido.Boffi@pv.infn.it}
\  and BARBARA PASQUINI\footnote{
Barbara.Pasquini@pv.infn.it}
}

\address{Dipartimento di Fisica Nucleare e Teorica, Universit\`a di Pavia\\
Via Bassi 6, I-27100 Pavia, Italy, and\\
Istituto Nazionale di Fisica Nucleare, Sezione di Pavia, Pavia, Italy\\
}

\maketitle

\pub{Received (Day Month Year)}{Revised (Day Month Year)}

\begin{abstract}
The parton content of the nucleon is explored within a meson-cloud model developed to derive light-cone wave functions for the physical nucleon. The model is here applied to study electromagnetic form factors, distribution amplitudes and nucleon-to-meson transition distribution amplitudes.

\end{abstract}

\ccode{PACS numbers: 12.39.Ki, 13.40Gp, 14.20.Dh}

%%%%%%%%%%%%%%%%%%%%%%%%%%%%%%%%%%%%%%%%%%%%%%%%%%%

\section{Introduction}	

Schematically, in light-front quantization a nucleon state is described by the following superposition
\be
\label{eq:fock}
\ket{\tilde N} = \psi_{(3q)}\ket{qqq} + \psi_{(3q+1g)}\ket {qqqg} +\psi_{(3q+q\bar q)} \ket{qqqq\bar q} +\dots,
\ee
where in the light-cone gauge $A^+=0$ the valence three-quark light-cone wave function (LCWF) $\psi_{(3q)}$ involves six independent amplitudes corresponding to different combinations of quark orbital angular momentum and helicity, and the Fock component $ \psi_{(3q+1g)}$ with three quarks plus one gluon involves 126 independent amplitudes.\cite{Ji:2002xn} Adding a pair of sea quarks into the valence component to build the amplitude $\psi_{(3q+q\bar q)}$ leads to an even more complicated LCWF. 

To probe the parton content of the nucleon suitable models have to be invented to give explicit expressions for the LCWFs. Here, LCWFs are derived in a meson-cloud model\cite{Pasquini:2006dv,Pasquini:2007iz} where the physical nucleon is pictured to be part of the time a bare nucleon and part of the time a baryon-meson system as a consequence of the spontaneously-broken chiral symmetry. Configurations are considered where the baryon can be a nucleon or a $\Delta$ and the meson can be a pion as well as a vector meson such as the $\rho$ or the $\omega$. The meson cloud will manifest itself as an extension of the charge distribution of protons and neutrons, which should be observable in the electromagnetic form factors at relatively small values of $Q^2$. The valence quark content of the LCWF will determine the distribution amplitudes (DAs) driving the soft contribution to the high $Q^2$ behavior of form factors. Further information on the LCWF can be obtained by studying transition distribution amplitudes (TDAs) occurring in the description, e.g.,  of hard exclusive backward pion electroproduction\cite{Lansberg:2007ec} or in associated production of a pion and a high-$Q^2$ dilepton pair in $p\bar p$ annihilation.\cite{Lansberg:2007se}

%%%%%%%%%%%%%%%%%%%%%%%%%%%%%%%%%%%%%%%%%%%%%%%%%%%

\section{Light-cone wave functions and the meson-cloud model of the nucleon}

Following Refs.~\refcite{Pasquini:2006dv,Pasquini:2007iz} and references therein, in the one-meson approximation of the meson-cloud model the quantum state of the physical nucleon ($\tilde N$), with four-momentum $p_N^\mu=(p^-_N,p^+_N,{\mathbf p}_{N\perp})\equiv(p^-_N,\tilde p_N)$ and helicity $\lambda$, can be written as 
\begin{eqnarray}
|\tilde p_N,\lambda;\tilde N\rangle
& &
=\sqrt{Z}|\tilde p_N,\lambda; N\rangle
+
\sum_{B,M}
\int \frac{{\rm d}y{\rm d}^2{\mathbf k}_{\perp}}{2(2\pi)^3}\,
\frac{1}{\sqrt{y(1-y)}}
\sum_{\lambda',\lambda''}
\phi_{\lambda'\lambda''}^{\lambda \,(N,BM)}(y,{\mathbf k}_\perp)\nonumber\\
& &\ {}\times
|yp^+_N,{\mathbf k}_{\perp}+y{\mathbf p}_{N\perp},\lambda';B\rangle\,
|(1-y)p^+_N,-{\mathbf k}_{\perp}+(1-y){\mathbf p}_{N\perp},\lambda'';M\rangle,\quad
\label{eq:14}
\end{eqnarray}
where the function 
$\phi_{\lambda'\lambda''}^{\lambda\,(N,BM)}(y,{\mathbf k}_\perp)$ is the probability amplitude to find a physical nucleon with helicity $\lambda$ in a state consisting of a  virtual baryon $B$ and a virtual meson $M$, with the baryon (meson) having helicity $\lambda'$ ($\lambda''$), longitudinal momentum fraction $y$ ($1-y$) and transverse momentum ${\mathbf k}_\perp$ ($-{\mathbf k}_\perp$). From the normalization condition of the nucleon state one obtains the following condition on the renormalization factor $Z$
\be
1=Z+\sum_{B,M} P_{BM/N},
\label{eq:norm}
\ee
with
\be
P_{BM/N}=
\int \frac{{\rm d}y{\rm d}^2{\mathbf k}_{\perp}}{2(2\pi)^3}\,
\sum_{\lambda',\lambda''}
|\phi^{1/2(N,BM)}_{\lambda'\lambda''}(y,{\mathbf k}_\perp)|^2.
\label{eq:prob}
\ee
Adopting a light-cone constituent quark model for the hadron states $\ket{{\tilde{p}_H},\lambda;H}$ of the bare nucleon ($N=3$) and baryon-meson ($N=5$) components, and using the minimal Fock-state wave function in the light-cone formalism, one has
\bea
\ket{{\tilde{p}_H},\lambda;H} 
& &
= \sum_{\tau_i,\lambda_i} \int 
\left(\prod_{i=1}^N \frac{{\rm d} x_i}{\sqrt{x_i}}\right) \delta\left(1-\sum_{i=1}^N x_i\right)
\left(\prod_{i=1}^N
\frac{{\rm d}^2{\mathbf k}_{i\perp}}{2(2\pi)^3}\right)\,2(2\pi)^3\,
\delta\left(\sum_{i=1}^N {\mathbf k}_{i\perp}\right)\nonumber\\
& &\quad {}\times
\Psi_\lambda^{H,[f]}(\{x_i,{\mathbf k}_{i\perp};\lambda_i,\tau_i\}_{i=1,...,N})
\prod_{i=1}^N
\ket{x_i p^+_H, \, {\mathbf p}_{i\perp},\lambda_i,\tau_i},
\label{eq:18}
\eea
where 
$\Psi_\lambda^{H,[f]}(\{x_i,{\mathbf k}_{i\perp};\lambda_i,\tau_i\}_{i=1,..,N})$ is the momentum LCWF  which gives the probability amplitude for finding in the hadron $N$ partons ($N=3$ and $N=2$ for the baryon and meson case, respectively) with momenta
 $(x_i p^+_H, {\mathbf p}_{i\perp}=\,{\mathbf k}_{i\perp} + x_i {\mathbf p}_{H\perp})$, 
and spin and isospin variables $\lambda_i$ and $\tau_i,$ respectively. 

The wave function $\Psi_\lambda^{H,[f]}$ can be obtained by transforming the ordinary equal-time (instant-form) wave function $\Psi_\lambda^{H,[c]}$ in the rest frame into that in the light-front dynamics, by taking into account relativistic effects such as the Melosh-Wigner rotation. The instant-form wave function is  here constructed as the product of a momentum wave function, which is spherically symmetric and invariant under permutations, and a spin-isospin wave function, which is uniquely determined by SU(6)-symmetry requirements.

In the case of the nucleon, the momentum wavefunction of Ref.~\refcite{Schlumpf:1992ce} has been adopted. It reads
\bea
\label{eq:psifc}
\psi^{N,[c]}(\{{\mathbf k}_i\}_{i=1,2,3})
=\frac{N'}{(M_0^2+\beta^2)^\gamma},
\label{eq:76}
\eea 
with $N'$  a normalization factor and $M_0=\sum_{i=1}^3 \omega_i=\sum_{i=1}^3\sqrt{m^2+\mathbf k_i^2}$ the mass of the noninteracting three-quark system. In Eq.~(\ref{eq:76}), the scale $\beta$, the parameter $\gamma$ for the power-law behaviour, and the quark mass $m$ are taken as free parameters. They have been fixed\cite{Pasquini:2007iz}  to fit the magnetic moments $\mu^p$ and $\mu^n$, the proton axial coupling constant $g_A=G_A(0)$, the Sachs form factors $G_E^n$ at $Q^2=0.15$ GeV$^2$, and $G_E^p$ and $G_M^p$  at $Q^2=0.15$ and  $0.45$ GeV$^2$. A 5\% uncertainty was allowed in the fitting procedure. A rather good fit is obtained in the proton case in the whole range of available data both for the Sachs form factors and the axial form factor $G_A$, while in the neutron case the fit is less satisfactory. In any case, the contribution from the meson cloud is smooth and mainly significant for $Q^2< 0.5$ GeV$^2$ with a maximum at $Q^2=0$. The fitted values are $\gamma=3.21$, $\beta=0.489$ GeV and $m=0.264$ GeV. These values differ from the original set of parameters ($\gamma=3.5$, $\beta=0.607$ GeV and $m=0.263$ GeV)\cite{Schlumpf:1992ce}, defined in order to reproduce $\mu^p$ and $\mu^n$ only.

The canonical wavefunction of the pion is taken from Ref.~\refcite{Choi:1997iq} and reads 
\begin{eqnarray}
\psi^{\pi,[c]}(\mathbf{k}_1, \mathbf{k}_2)
=\frac{i}{\pi^{3/4}\alpha^{3/2}}
\exp{[-k^2/(2\alpha^2)]},
\label{eq:can_psi}
\end{eqnarray}
with $\mathbf k= \mathbf k_1=-\mathbf k_2$, and the two parameters  $\alpha=0.3659$ GeV and $m_q=0.22$ GeV 
fitted to the pion form factor data. The wavefunction of the $\rho$ differs from the pion only in the spin 
component, with the rest-frame spin states of the $q\bar q$ pair coupled to $J=1$ instead of $J=0$.
Similarly, the $\omega$ is described by the same spin and momentum wavefunction as the $\rho$, 
but with the isospin component corresponding to a singlet state.

With such ingredients, the probabilities of the bare nucleon state, the ($p\pi$) and ($\Delta\pi$) configurations are 90\%, 5.1\% and 3.4\%, respectively. 

%%%%%%%%%%%%%%%%%%%%%%%%%%%%%%%%%%%%%%%%%%%%%%%%%%%

%++++++++++++++
\begin{figure}[pb]
\centerline{\psfig{file=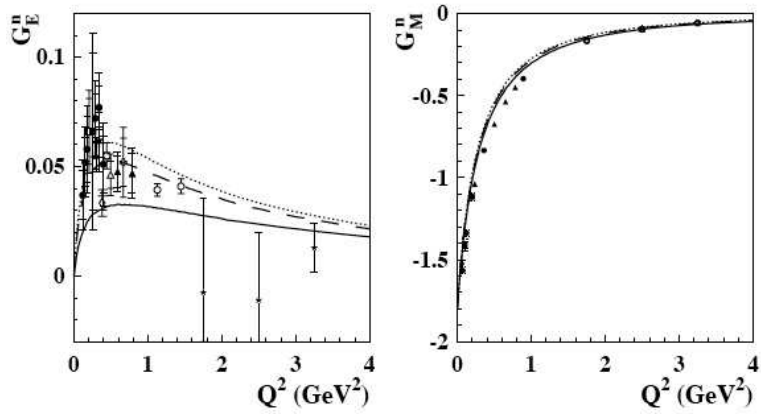,width=4in}}
\vspace*{8pt}
\caption{The electric form factor of the neutron. Solid line for the bare neutron wavefunction with SU(6) symmetry, dashed (dotted) line including 1\% (2\%) mixed-symmetry  $S$-state.}
\label{fig:fig1}
\end{figure}
%++++++++++++++

\section{Nucleon charge and magnetization densities}

It is known that the presence of a small admixture (1--2\%) of mixed-symmetry $S$-wave components greatly improves the result for the neutron electric form factor. Following Ref.~\refcite{JuliaDiaz:2003gq} the mixed-symmetry $S$-wave component is assumed to be represented by an appropriate combination of mixed-symmetry spin-isospin wavefunctions with two radial wavefunctions of mixed symmetry. The consequences of including a small percentage of such mixed-symmetric contribution to the neutron electric form factor are illustrated in Fig.~\ref{fig:fig1}. Even a percentage as small as 1\% is able to produce a quite good result compared to data. In particular, including also a mixed-symmetry $S$-wave component the neutron charge radius becomes quite close to the experimental value ($r^2_n=-0.112$ fm$^2$). As anticipated in Ref.~\refcite{JuliaDiaz:2003gq} the same calculation leaves the other nucleon form factors almost unaffected. 

%++++++++++++++
\begin{figure}[pb]
\centerline{\psfig{file=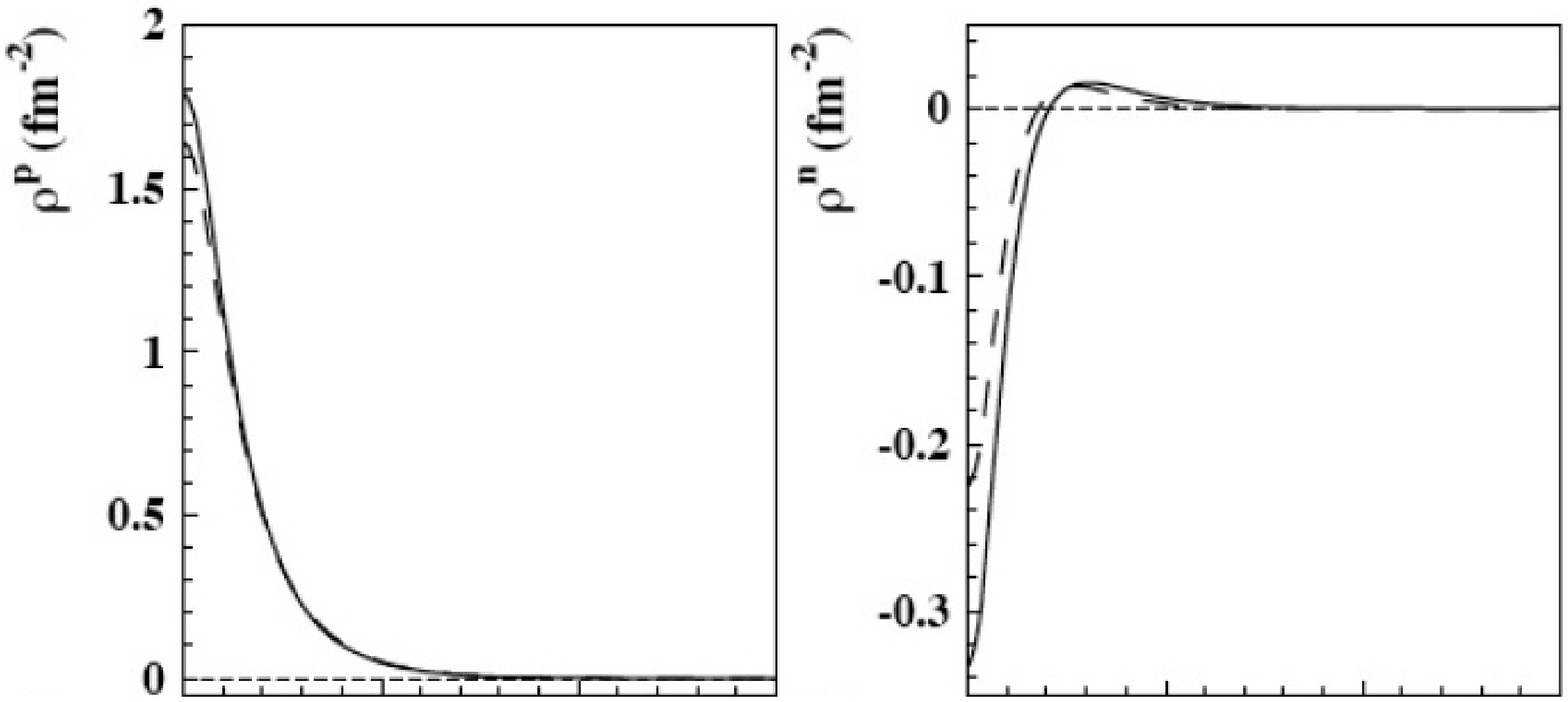,width=3.8in}}
\centerline{\psfig{file=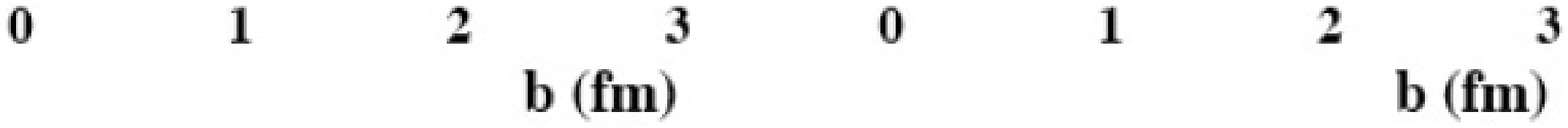,width=4in}}
\vspace*{8pt}
\caption{The proton and neutron charge densities as a function of the impact parameter $b$. Solid lines for a permutationally symmetric momentum wave function, dashed lines including mixed-symmetry components.}
\label{fig:fig2}
\end{figure}
%++++++++++++++

The radial distributions of the nucleon charge ($\rho_{ch}$) and magnetization ($\rho_m$) have been estimated for a long time by taking the Fourier-Bessel transform of the Sachs form factors in the Breit frame and neglecting relativistic corrections. The problem of unambiguously determining the charge density can be solved by looking at the charge density $\rho(b)$ of partons in the transverse (impact parameter) plane as a function of the distance $b$ relative to the transverse center of longitudinal momentum.\cite{Miller:2007uy} This is possible because in the transverse plane boosts are purely kinematical, i.e. in the light-front framework they form a Galilei subgroup of the Poincar\'e group. Then $\rho(b)$ is the two-dimensional Fourier transform of the Dirac form factor $F_1$:
\be
\label{eq:rhodib}
\rho(b) = \frac{1}{2\pi}\int_0^\infty {\rm d}Q\,Q \,J_0(Qb) F_1(Q^2),
\ee
where $J_0 $ is a cylindrical Bessel function. 

The corresponding charge densities for the proton and the neutron are plotted in Fig.~\ref{fig:fig2}. They are concentrated at low values of $b$ with a positive peak for the proton and a negative peak for the neutron, in agreement with the phenomenological analysis of Ref.~\refcite{Miller:2007uy}.

%++++++++++++++
\begin{figure}[pb]
\centerline{\psfig{file=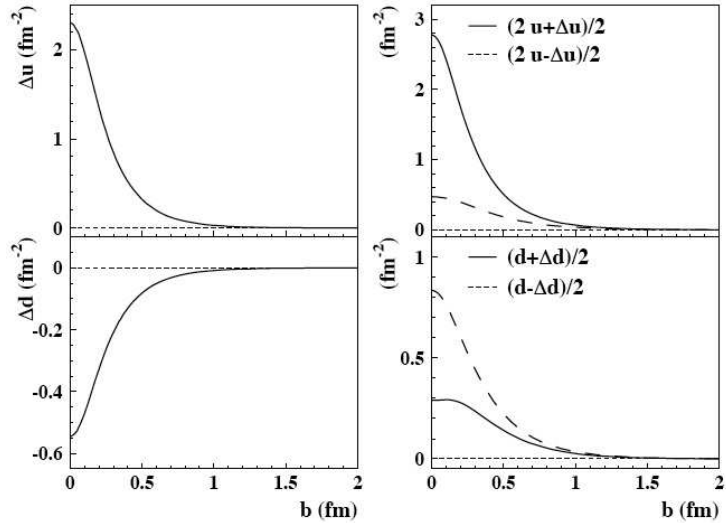,width=3.8in}}
%\vspace*{8pt}
\caption{Transverse distribution of up and down quarks in a longitudinally polarized proton as a function of the impact parameter $b$. Left panels: the axial contributions $\Delta u$ and $\Delta d$ for up and down quarks, respectively. Right panels: total contribution for quarks polarized in the longitudinal direction, either parallel (solid lines) or antiparallel (dashed lines) to the proton helicity.}
\label{fig:fig3}
\end{figure}
%++++++++++++++

These nucleon charge densities can be related to quark transverse distributions. Assuming that only up and down quarks are in the nucleon and invoking isospin symmetry, $u(b)$ and $d(b)$ can be obtained using
\be
u(b) =  \rho^p(b) +\frac{1}{2} \rho^n(b), \quad
d(b) = \rho^p(b) + 2\rho^n(b).
\ee
Making additional use of the axial form factor, one can derive the probability $\rho^q(b,\lambda,\Lambda)$ to find a quark with transverse position $b$ and light-cone helicity $\lambda$ ($=\pm 1$) in the nucleon with longitudinal polarization $\Lambda$ ($=\pm 1$),\cite{Pasquini:2007iz} i.e.
\be
\rho^q(b,\lambda,\Lambda) =  \frac{1}{2}\left[\rho^q(b) +  \lambda\Lambda \Delta q(b)\right],
\ee
where $\rho^q(b)$ is defined in Eq.~(\ref{eq:rhodib}) and $\Delta q(b)$ is the Fourier transform of $G^q_A(Q^2)$. Assuming a positive proton helicity ($\Lambda=1$) the resulting probability is shown in the right panels of Fig.~\ref{fig:fig3}. The axial contributions $\Delta u (b)$ and $\Delta d (b)$ for up and down quarks (left panels), respectively, have opposite sign. When suitably combined with the corresponding transverse distributions $u(b)$ and $d(b)$ the positive helicity up quarks in the proton turn out to be preferentially aligned with the proton helicity, while the opposite occurs for down quarks. This result corresponds to that shown in  Fig. 7 of Ref.~\refcite{Pasquini:2007xz} where quite a different radial distribution of the axially symmetric spin density was presented for up and down quarks in  the transverse plane.

%%%%%%%%%%%%%%%%%%%%%%%%%%%%%%%%%%%%%%%%%%%%%%%%%%%

\section{Distribution amplitudes}

The proton DAs are derived from the following proton-to-vacuum matrix elements of trilocal operators built of quark and gluon fields\cite{Chernyak:1984bm}
\begin{equation}
\label{General-Matrix-Element}
D^\lambda_{\alpha,\beta,\gamma} = 
\mathcal{F}\left(\langle 0|\epsilon^{ijk}u^{i'}_\alpha(z_1n)[z_1;z_0]_{i'i}
u^{j'}_\beta(z_2n)[z_2;z_0]_{j'j}
d^{k'}_\gamma(z_3n)[z_3;z_0]_{k'k}|p(p_p,\lambda)\rangle\right),
\end{equation}
where ${\cal F}$ means Fourier transform, $\ket{p(p_p,\lambda)}$ denotes the proton state with momentum $p_p$ ($p_p^2=M^2$) and helicity $\lambda$; $u$, $d$ are the field operators for up and down quarks, respectively; the Greek letters $\alpha$, $\beta$ and $\gamma$ stand for Dirac indices, while the Latin letters $i$, $j$ and $k$ refer to color; $n$ is an arbitrary light-like vector ($n^2=0$) and $z_i$ are real numbers that specify quark separation, with $\sum_iz_i=1$. In Eq.~(\ref{General-Matrix-Element}) the gauge factors $[z_i ; z_0]$ render the matrix element gauge-invariant; in the following the light-cone gauge  $A^+ = 0$ will be adopted, so that the gauge factors reduce to the identity.

%++++++++++++++
\begin{figure}[pb]
\centerline{\psfig{file=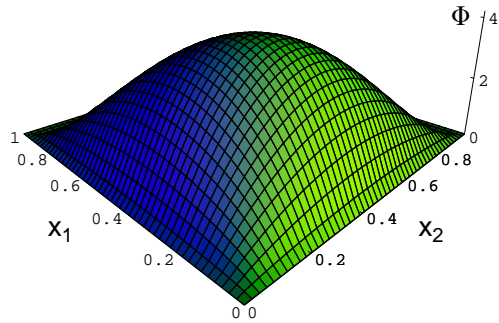,width=16pc}}
\vspace*{8pt}
\caption{The proton distribution amplitude.}
\label{fig:figure1_DAs}
\end{figure}
%++++++++++++++

%++++++++++++++
\begin{table}
%\begin{center}
\tbl{The moments $\phi^{(l,m,n)}$ of the proton DA at different scales. The asymptotic values (AS) 
and the lattice results (LAT) at the scale $Q^2=1$ GeV$^2$ are compared with the corresponding results from the present model calculation (Ref. 12), after evolution to $Q^2=1$ GeV$^2$.}
{\begin{tabular}{@{}ccll@{}}
\toprule
($l$,\,$m$,\,$n$)   
&
AS
&
\qquad LAT
&
Ref. 12\\ 
\colrule
$(0\;\; 0\;\; 0)\,$ & $\,    1    \,$                        & $\,   1    \,$ & $\,   1    \,$ \\
$(1\;\; 0\;\; 0)\,$ & $\,\onet\simeq 0.333\,$ & $\,     0.3999(37)(139)   \,$  &  $\, 0.340\,$  \\
$(0\;\; 1\;\; 0)\,$ & $ \onet\simeq 0.333$ & $\,   0.2986(11)(52)    \,$ &  $\, 0.335\,$ \\
$(0\;\; 0\;\; 1)\,$ & $\onet\simeq 0.333$& $\, 0.3015(32)(106) \,$ &$\, 0.326\,$\\
$(2\;\; 0\;\; 0)\,$ & $\oneseven\simeq 0.143$ & $\, 0.1816(64)(212) \,$ &$\, 0.147\,$\\
$(0\;\; 2\;\; 0)\,$ & $\oneseven\simeq 0.143$ & $\,  0.1281(32)(106)  \,$ &$\, 0.144\,$\\
$(0\;\; 0\;\; 2)\,$ & $\oneseven\simeq 0.143$& $\,  0.1311(113)(382)  \,$ &$\,0.137\,$ \\
$(1\;\; 1\;\; 0)\,$ & ${\textstyle {2\over 21}}\simeq 0.095$ & $\, 0.1092(67)(219)  \,$ &$\, 0.098\,$\\
$(1\;\; 0\;\; 1)\,$ & ${\textstyle {2\over 21}}\simeq 0.095$ & $\,  0.1091(41)(152)  \,$ &$\, 0.095\,$\\
$(0\;\; 1\;\; 1)\,$ & ${\textstyle {2\over 21}}\simeq 0.095$ & $\,  0.0613(89)(319)  \,$ &$\,0.093\,$ \\
\botrule
\end{tabular}\label{Moments_Evol}}
%\end{center}
\end{table}
%++++++++++++++
The matrix elements $D^\lambda_{\alpha\beta,\gamma}$, and ultimately the DAs, are directly linked to the $L_z=0$ component of the valence-quark wave function of the nucleon, by integrating out the transverse momenta of the constituent quarks. To the leading twist-three accuracy this amounts to writing the three-quark $uud$ component of the proton state with positive helicity in the infinite momentum frame in terms of a single independent scalar function $\Phi(x_1,x_2,x_3)$, where the variables $x_i$, conjugate to the light-cone positions of the quark operators, are collinear momentum fractions  of the proton longitudinal momentum carried by each quark, with $0\leq x_i\leq 1$ and $\sum_{i=1}^3x_i=1$ by momentum conservation. One has\cite{Pasquini:2009ki}
\bea
\ket{p(p_p,\uparrow)}_{uud}
&=&\frac{1}{\sqrt{3}} \frac{f_N}{4}
\int_0^1\left(\prod_{i=1}^3 \frac{{\rm d} x_i}{\sqrt{x_i}}\right) \delta\left(1-\sum_{i=1}^3 x_i\right)
\Phi(x_1,x_2,x_3)\nonumber\\
&&\quad {}\times
\left[\ket{u^\uparrow(\tilde k_1)u^\downarrow(\tilde k_2)d^\uparrow(\tilde k_3)}-\ket{u^\uparrow(\tilde k_1)d^\downarrow(\tilde k_2)u^\uparrow(\tilde k_3)} \right],
\label{DA-proton-state2}
\eea
where $f_N$ is the proton decay constant.

Correspondingly, with the LCWFs of the previous section one obtains
\bea
\Phi(x_1,x_2,x_3) 
&=& -
\frac{8\sqrt{3}}{f_N} \int \left(\prod_{i=1}^3
\frac{{\rm d}^2{\mathbf k}_{i\perp}}{2(2\pi)^3}\right)\,2(2\pi)^3\,
\delta\left(\sum_{i=1}^3 {\mathbf k}_{i\perp}\right)\nonumber\\
&&\quad{}\times
\frac{2(2\pi)^3}{\sqrt{M_0}} \left[ \prod_{i=1}^3 \left( \frac{\omega_i}{x_i}\right)^{1/2}\right]
\psi(\{x_i, \mathbf{k}_{i\perp}\})
\Xi_\uparrow^{p}(\uparrow, \downarrow,\uparrow),
\eea
where $\psi(\{x_i, \mathbf{k}_{i\perp}\})$ is given by Eq.~(\ref{eq:76}) and
\be
 \Xi^{p}_{\uparrow}\left(\uparrow,\downarrow,\uparrow\right)
=\frac{1}{\sqrt{6}}\prod_i\frac{1}{\sqrt{N(k_i)}}
\,(-a_1a_2a_3+k_1^L k_2^R a_3-2a_1k_2^Rk_3^L)
\ee
with the following definitions:
$a_i=(m+x_i M_0)$, $N(k_i)=[(m+x_i M_0)^2+k^2_{i\perp}]$,
$k_i^{R}=k_{i\,x}+ik_{i\,y}$, and  $k_i^{L}=k_{i\,x}-ik_{i\,y}$.

In Fig.~\ref{fig:figure1_DAs} the model results for the proton distribution amplitude $\Phi$ are shown. The resulting shape is quite similar to that of the symmetric asymptotic DA $\Phi_{as}=120x_1x_2x_3$. 
The results for the first and second moments $\phi^{(l,m,n)}$ of $\Phi$ with $l+m+n\leq 2$ after evolution to $Q^2=1$ GeV$^2$ are shown in Table~\ref{Moments_Evol}. Comparison with the analogous results from lattice estimates\cite{Gockeler:2008xv} is quite nice. 

%%%%%%%%%%%%%%%%%%%%%%%%%%%%%%%%%%%%%%%%%%%%%%%%%%%

\section{Transition distribution amplitudes}

%++++++++++++++
\begin{figure}[pb]
 	\centering
\epsfig{file=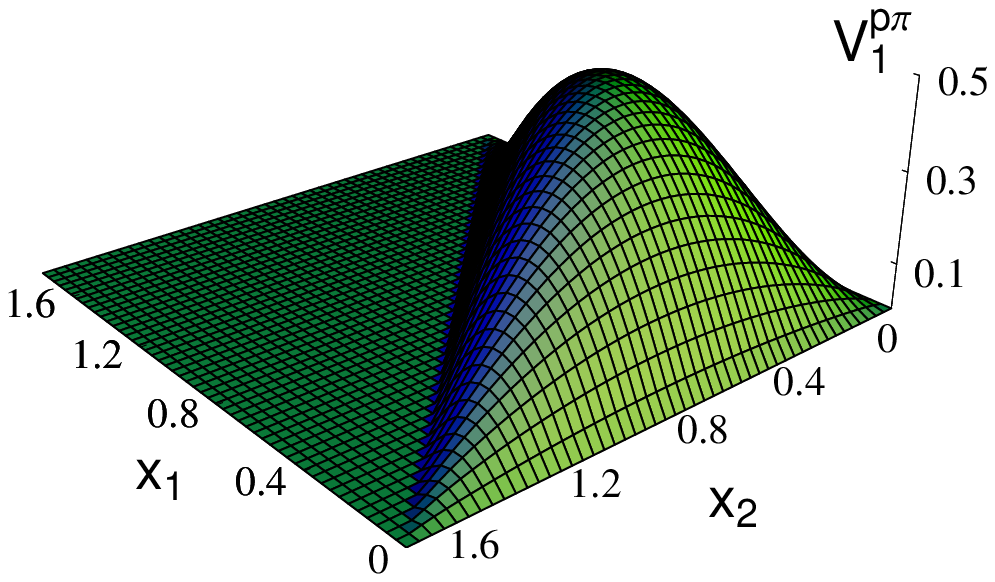,  width=14 pc}
\hspace{0.5 truecm}
\epsfig{file=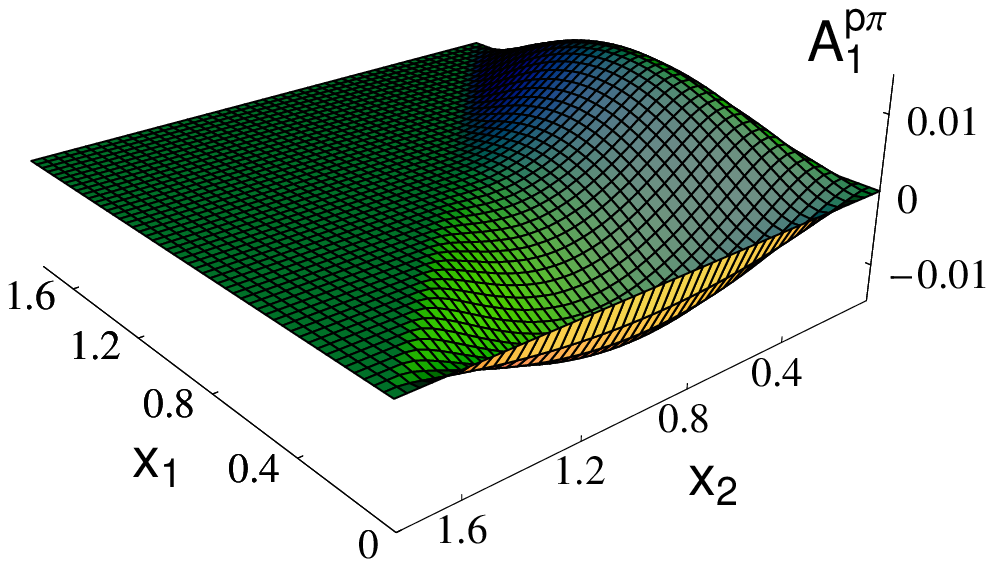,  width=14 pc}
\epsfig{file=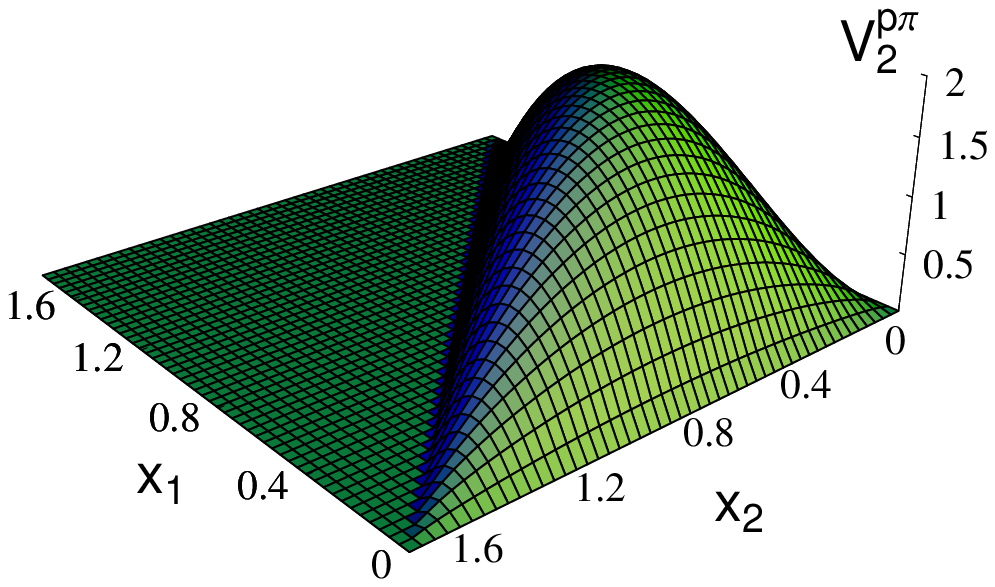,  width=14 pc}
\hspace{0.5 truecm}
\epsfig{file=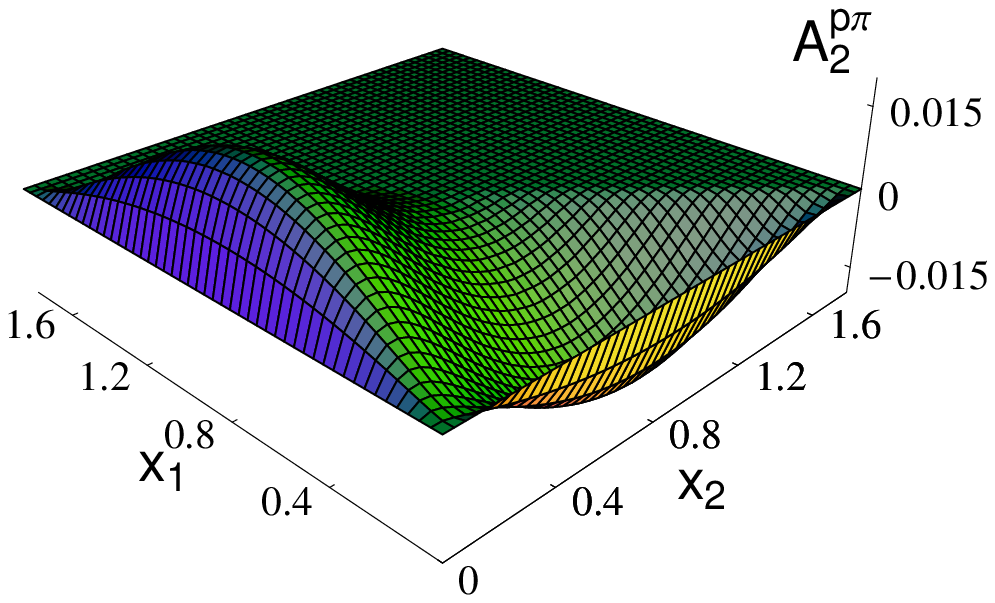,  width=14 pc}
\caption{The $p\rightarrow \pi^0$  TDAs  $V^{p\pi}_1$ (up left), $A^{p\pi}_1$ (up right), $V^{p\pi}_2$ (down left), $A^{p\pi}_2$ (down right) as functions of $(x_1,x_2,2\xi-x_1-x_2)$ at fixed $\xi=0.9$ and $\Delta^2=-0.1$ GeV$^2$. }
\label{f6}
\end{figure}
%++++++++++++++

%++++++++++++++
\begin{figure}[t]
 	\centering
\epsfig{file=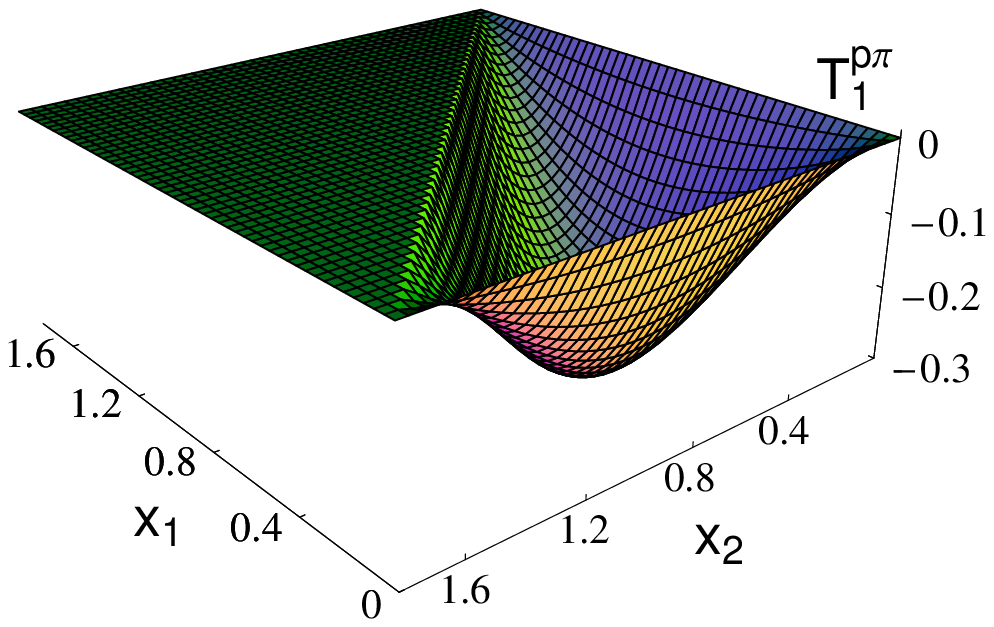,  width=14 pc}
\hspace{0.5 truecm}
\epsfig{file=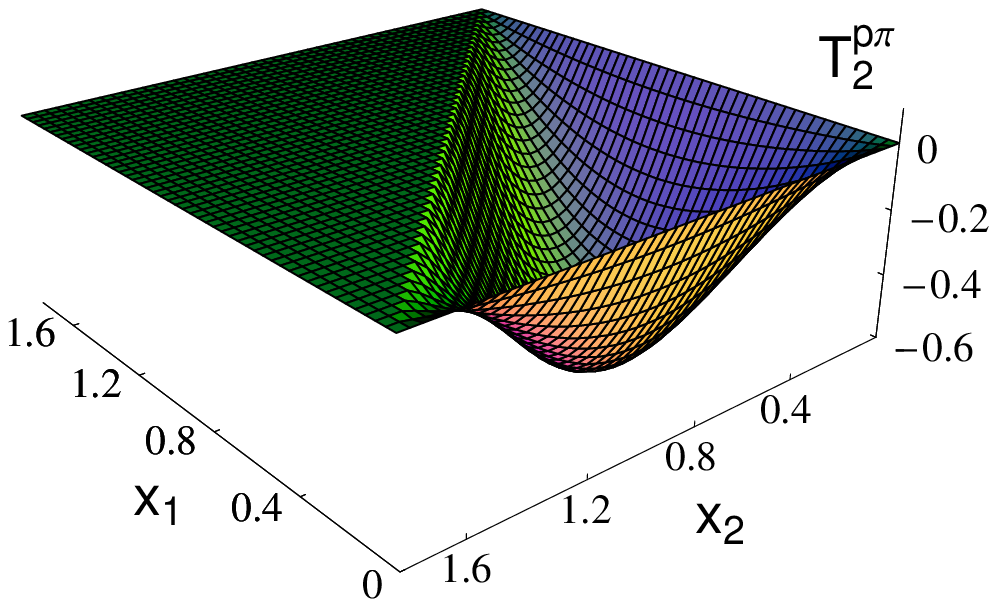,  width=14 pc}
\epsfig{file=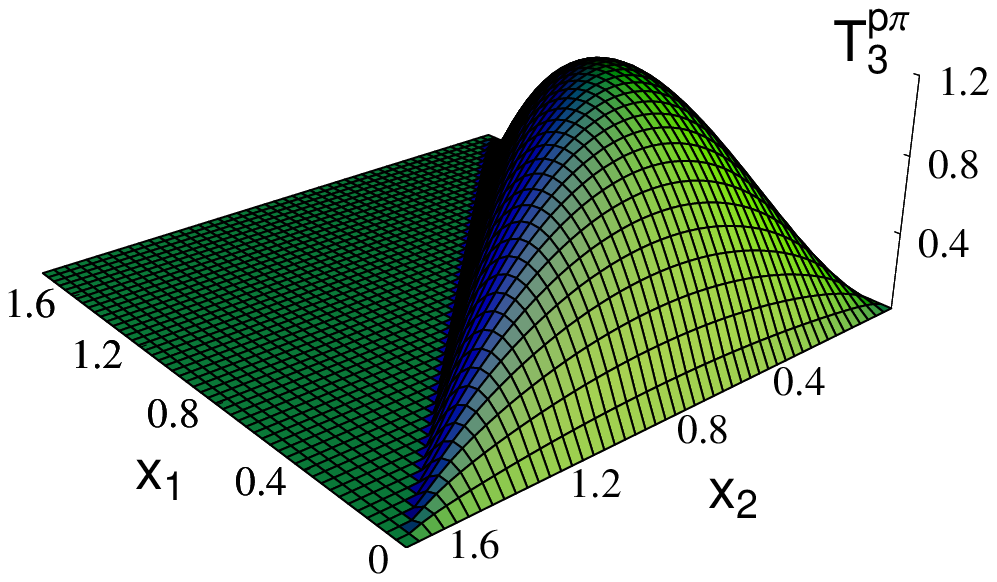,  width=14 pc}
\hspace{0.5 truecm}
\epsfig{file=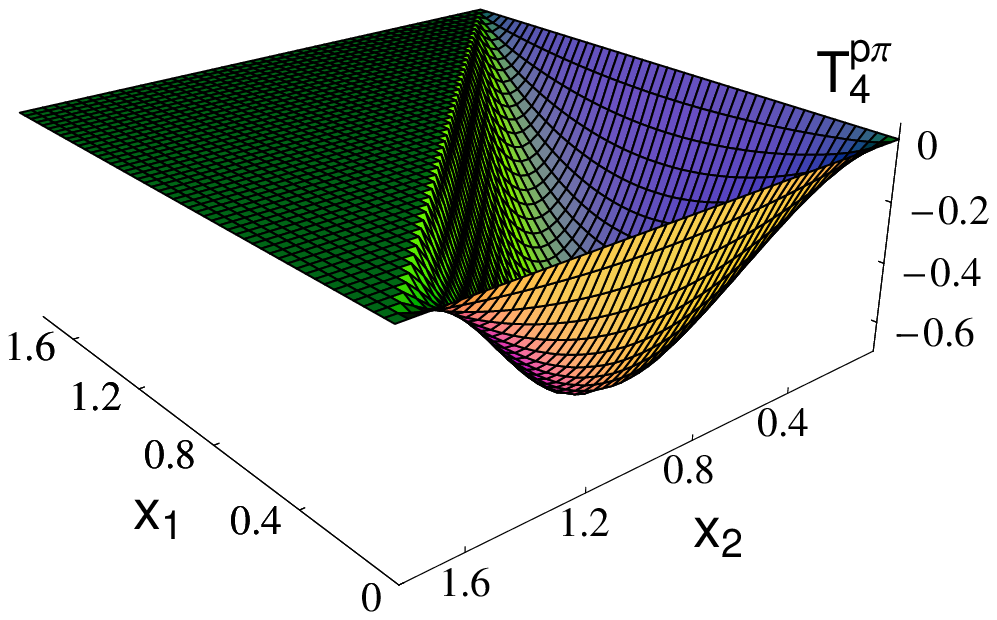,  width=14 pc}
\caption{The  $p\rightarrow \pi^0$ TDAs $T^{p\pi}_1$ (up left), $T^{p\pi}_2$ (up right), $T^{p\pi}_3$ (down left), and $T^{p\pi}_4$ (down left) as functions of $(x_1,x_2,2\xi-x_1-x_2)$ at fixed $\xi=0.9$ and $\Delta^2=-0.1$ GeV$^2$.}
\label{f7}
\end{figure}
%++++++++++++++

The leading-twist decomposition of the general matrix element describing the transition from a nucleon to a meson state, e.g. a $\pi^0$,
\be
T^\lambda_{\alpha,\beta,\gamma} =
4\mathcal{F}\left(\langle\pi^0(p_\pi)|\epsilon^{ijk}u^i_\alpha(z_1n)
u^j_\beta(z_2n) d^k_\gamma(z_3n)|p(p_p,\lambda)\rangle\right) ,
\ee
involves eight TDAs.\cite{Pire:2005ax} The TDAs are dimensionless functions and depend on $(x_i,\xi,\Delta^2)$,
where the fraction of quark plus momentum $x_i$ have support in $[-1+\xi,1+\xi]$, 
\be
\Delta^2= (p_p - p_\pi)^2 = -2\xi\left[\frac{m_\pi^2+\Delta_\perp^2}{1-\xi}
-\frac{M^2}{1+\xi}\right]-\Delta_\perp^2,
\ee
and the skewness variable $\xi$ describes the loss of plus momentum of the initial hadron in the proton-to-meson transition, i.e.
\be
\xi = -\frac{\Delta\cdot n}{2P\cdot n}=-\frac{\Delta^+}{2P^+}, \quad
\mbox{with}\;\;
P =\frac{1}{2}(p_p+ p_\pi).
\ee
Restricting ourselves to the case $\xi>0$, momentum conservation requires $\sum_ix_i=2\xi$. The fields with positive momentum fractions, $x_i\geq 0$, describe creation of quarks, whereas those with negative fractions, $x_i\leq 0$, the absorption of antiquarks. This leads to define three distinct kinematical regions: the ERBL region for $x_i\geq 0$, and two DGLAP regions when $x_1\geq 0$, $x_2\geq 0$, $x_3\leq 0$,  or $x_1\geq 0$, $x_2\leq 0$, $x_3\leq 0$. The names derive from the evolution equations which control the scale dependence of the TDAs in the different regions.\cite{Pire:2005ax} 

Results are presented in Figs. \ref{f6} and \ref{f7} for the TDAs in the ERBL region, corresponding to probe the $\psi_{3q+q\bar q}$ and the contribution from the fluctuations of the nucleon in ($p\pi^0$) and ($\Delta^+\pi^0$) subsystems. The relative contribution of these components depends on the momentum transferred between the initial nucleon and the final pion as well as on the different spin configurations of the intermediate baryon. In particular, the $\Delta$ plays a special role in the case of the tensor TDAs $T^{p\pi}_i$ which involve configurations with helicity $\pm3/2$ (the only nonvanishing contribution to $T^{p\pi}_4$), while the interplay of the nucleon and $\Delta$ contributions with helicity $\pm1/2$ determines the different shape of the vector ($V^{p\pi}_i$) and axial-vector ($A^{p\pi}_i$) TDAs.\cite{Pasquini:2009ki}
 
%%%%%%%%%%%%%%%%%%%%%%%%%%%%%%%%%%%%%%%%%%%%%%%%%%%

\section{Concluding remarks}

In the light-cone description the nucleon state is decomposed in terms of $N$-parton Fock states with coefficients representing momentum light-cone wave functions of the $N$ partons. Since the constituent quark models work so well phenomenologically, in applications it is usually assumed that only the Fock components with a few partons have to be taken into account. One of such models has been studied in a series of papers to show that the parametrization of the LCWFs up to five-parton components is already sufficient to account for the electroweak form factors\cite{Pasquini:2007iz} and spin densities\cite{Pasquini:2007xz} of the nucleon, as well as the observed asymmetries due to transverse momentum dependence of parton distributions,\cite{Pasquini:2008ax} and to give a useful insight into the quark generalized parton distributions.\cite{Boffi:2002yy,Boffi:2007yc,Pasquini:2005dk,Pasquini:2006dv} As a further test of the model in this paper the nucleon distribution amplitudes and the nucleon-to-meson transition distribution amplitudes have been considered. 

%%%%%%%%%%%%%%%%%%%%%%%%%%%%%%%%%%%%%%%%%%%%%%%%%%%

\section*{Acknowledgments}

%It is a pleasure to acknowledge a fruitful collaboration with Barbara Pasquini with whom most of the results presented in this paper have been obtained. The author is also gratefully indebted to  the organizers for the invitation to the Workshop on Recent Advances in Perturbative QCD and Hadronic Physics in honor of Professor  A. V. Efremov to whom he also expresses the best wishes. 
It is pleasure for the authors to take the opportunity of the kind invitation to the Workshop on Recent Advances in Perturbative QCD and Hadronic Physics in honor of Professor  A. V. Efremov to express to him the best wishes.
This work is part of %the Research Infrastructure Integrating Activity ``Study of Strongly Interacting Matter'' (acronym the activity HadronPhysics2, Grant Agreement n. 227431) 
the activity HadronPhysics2, Grant Agreement n. 227431, under the Seventh Framework Programme 
of the European Community.

%%%%%%%%%%%%%%%%%%%%%%%%%%%%%%%%%%%%%%%%%%%%%%%%%%%

\end{document}